# The Pioneer Anomaly and a Rotating Gödel Universe

T. L. Wilson[a] and H.-J. Blome[b]

[a]National Aeronautics and Space Administration, 2101 NASA Parkway, Houston, TX 77058 USA
[b]University of Applied Sciences, Hohenstaufenallee 6, 52064 Aachen, Germany

**Abstract**

Based upon a simple cosmological model with no expansion, we find that the rotational terms appearing in the Gödel universe are too small to explain the Pioneer anomaly. Following a brief summary of the anomaly, cosmological effects on the dynamics of local systems are addressed - including a derivation of the equations of motion for an accelerated Pioneer-type observer in a rotating universe. The rotation or vorticity present in such a cosmological model is then subjected to astrophysical limits set by observations of the cosmic microwave background radiation. Although it contributes, universal rotation is not the cause of the Pioneer effect. In view of the related fly-by anomalies, frame-dragging is also discussed. The virial theorem is used to demonstrate the non-conservation of energy during transfers from bound to hyperbolic trajectories.

## 1. Introduction

A number of studies investigating the radio and Doppler ranging data measured for Pioneer 10 and 11 found the presence of a small unexpected blue frequency shift when these spacecraft began exploring the outer solar system beyond 20 AU (Anderson et al. 2002, 1998; Turyshev 2006, 2005; Nieto, 2004). If not a systematic error, the drift can be interpreted as due to a constant deceleration of $a_p = \sim 8.74 \times 10^{-8}$ cm s$^{-2}$ back in the direction of the Sun. Now known as the Pioneer anomaly, it is important to note that the shift is constant and is the same for two spacecraft travelling in virtually opposite directions away from the Sun. A possibly related effect has also been noted during Earth fly-by's that use osculating orbits and patched conics for gravity assists, the so-called fly-by anomaly (Anderson et al. 2008; Lämmerzahl et al. 2006) which affects velocity rather than acceleration. Among the many conjectures as to the cause of these anomalies, rotational dynamics seems to be the most seriously discussed subject. For the Pioneer anomaly per se, several new aspects of rotational dynamics involving the possibility of a rotating Gödel universe (Gödel 1949) and frame dragging in relativity for hyperbolic trajectories will be addressed here.

## 2. Summary of the Pioneer Anomaly

Any physical object in space is subject to numerous kinds of forces, gravitational and non-gravitational. Therefore, a spacecraft in the Solar System experiences a complex environment of dynamic accelerations that alter both its rotational attitude as well as its translational trajectory. Gravitational effects include perturbations by unknown objects in the Kuiper Belt (Nieto 2005), unmeasured harmonics in gravitational potentials, nonlocal effects in the 3-body problem for gravity assists during fly-by's, the transition to hyperbolic orbits such as Pioneer 11 at 9.39 AU near Saturn, and possible changes in local dynamics produced by cosmological expansion. Non-gravitational forces include the solar wind, Poynting-Robertson drag (wrong sign), drag produced by the interplanetary medium such as dust and neutrals from the local interstellar medium, dark matter drag if it exists, spacecraft maneuvers, solar corona modelling, anisotropic system thermal radiation, and systematics involving heat and thruster gas leaks. Adding noise, uncertainties, and statistical filtering of data, all of these must be modelled and then shown to fit the Pioneer flight data. To date, they do not.

It was only after long and careful modelling of most of the obvious forces acting on Pioneer that the acceleration anomaly arose and was disclosed (Anderson et al. 2002, 1998). Lengthy details are in the literature and further analysis of additional data is proposed. Nieto (2008) has pointed out that more care needs to be taken when searching for the onset of the anomaly as related to the trajectory transition from a bound conic to a hyperbolic orbit at Saturn. He has proposed that the New Horizon mission currently on its way to Pluto and the Kuiper Belt can contribute valuable information about onset of the anomaly that may eliminate several of the systematic uncertainties. New Horizons will also add a third spacecraft to the data set. It is even possible that this mission will not experience the Pioneer-like anomaly as has been the case where some fly-by's such as Messenger at Mercury (Anderson et al. 2008) do not see a fly-by velocity anomaly.

## 3. Cosmological Effects On Local Systems

The subject of cosmological effects on local systems is as old as relativistic cosmology itself. Following the discovery of the expansion of the Universe, the question arose whether universal expansion necessarily meant that bound systems like the Solar System within it would also expand due to Hubble expansion of spacetime (e.g. McVittie 1933). The consensus of opinion has been negative, namely that the expansion causes miniscule changes in local dynamics that are too small to be measured if they occur at all. Another opinion is that local cosmological effects are in fact not measurable quantities – a notion that follows from quantum physics. Hence, a metaphor for students of cosmology has been to imagine the Universe as a raisin cake which rises but the raisins do not expand with it; or an expanding balloon with coins representing galaxies pasted to it. The coins do not expand even though the coins move apart as the balloon expands.

As a further example, the cosmological constant $\Lambda$ in General Relativity can be used to model expansion of the accelerating Universe in cosmology (Blome and Wilson 2005). The term also appears in the precession of planetary orbits predicted by Einstein's theory of gravitation. Although mathematically present, the miniscule $\Lambda$-term is simply not measurable by today's understanding of Keplerian conic physics.

Nevertheless, the question remains. Do atomic and planetary orbits change with universal expansion or not? Cooperstock et al. (1998), Adkins et al. (2007), Anderson (1995), and others (Mena et al. 2002; Bonnor 1999, 1996, 1987) have addressed this subject with renewed interest. In spite of the outcome of the Pioneer anomaly as an issue, it is clear that the failure to explain such an experimental observation has inspired an interesting discussion in theoretical physics regarding several long-standing questions.

### 3.1 The Pioneer Anomaly and Cosmology

Obviously, large-scale features of the Universe may have some bearing upon the local behavior of observable physical systems. This was Mach's argument as to the origin of local inertia. Similarly, if the Universe is rotating, that will also manifest itself at some level of local physics. Ironically, a rotating universe was shown by Gödel (1949) to dispel Einstein's argument that General Relativity is consistent with Mach's Principle for the origin of inertia. Nevertheless, the discussion continues.

To investigate the consequences of cosmic rotation on bound and unbound motion of a spacecraft such as Pioneer within the Solar System, embedded in a Gödel universe, one must estimate the forces acting on the spacecraft. These are of two principal types: (a) Those that arise from nongravitational acceleration and rotating-frame effects; and (b) those that derive from the Riemann curvature tensor in relativity. There exists a third category involving an admixture of the first two but these are very small and will not be considered here. Many research groups have addressed this broad subject in different ways.

The procedure is to set up a local inertial frame and derive the equations of motion from the two sources acting on a parallel-transported tetrad representing the observer's local coordinate system. However, that tetrad must also be allowed to rotate. As a consequence, two types of dynamics occur. (a) Rotation of the tetrad induces Coriolis and centrifugal or centripetal forces; and (b) Fermi-type parallel transport (Fermi 1922; Manasse and Misner 1963) of the tetrad results in the derivation of Riemann curvature contributions to some order of approximation. The resulting equations of motion then describe the acceleration experienced by an object passing through a universe without any restrictions as to expansion or rotation.

For the case of a nonrotating universe, Cooperstock et al. (1998) and Adkins et al. (2007) have already examined a first-order perturbation. The goal here is not to say that cosmic rotation causes the Pioneer anomaly, but rather to place limits on cosmic interaction with local observables in a rotating universe.

### 3.2 Equations of Motion for an Accelerated, Rotating Observer

The equations of motion for an accelerated, rotating laboratory in a spacetime without curvature follow from classical Hamilton-Jacobi theory and encompass Newtonian acceleration along with Coriolis and centrifugal (centripetal) accelerations. To these must be added the relativistic effects for Lorentz transformations that boost them to arbitrary velocities. And finally, gravitational accelerations must be introduced in the form of the Riemann curvature tensor as predicted by Einstein's theory of General Relativity for Fermi transport (Fermi 1922). Because General Relativity has Newton's theory as its classical limit, the results are internally consistent. Hence the collective process can be expressed in terms of differential Riemann geometry. The procedure has been addressed in several places (e.g. Ni and Zimmermann 1978; Li and Ni 1979).

For a free-falling observer in a weak gravitational background field, one sets up the Fermi frame (Riemann normal coordinates) to study the Riemann curvature tensor by using geodesic deviation equations. As mentioned, one must also require that the observer's frame rotates, a condition that is disallowed by Fermi transport. The result is a coordinate system that better corresponds to accelerations experienced by physical observers (Figure 1). This procedure is similar to and related to the method of Eddington-Robertson parameters and the post-Newtonian weak-field approximations that are used in studying the field equations of General Relativity.

The equations of motion follow from the geodesic deviation equation which measures the change in separation and hence relative acceleration between neighboring geodesics in a Riemann geometry,

$$\frac{dx^\mu}{d\tau^2} + \Gamma^\mu_{\alpha\beta}\frac{dx^\alpha}{d\tau}\frac{dx^\beta}{d\tau} = 0 \; , \quad (1)$$

where $x^\mu = (x^o, x^i) = (ct, \boldsymbol{x})$ is a spacetime 4-vector and boldface $\boldsymbol{x} = \vec{x} = (x^i, x^j, x^k)$ is a 3-vector with units $c=1$ for the speed of light. Greek indices vary from 0 to 4, Latin indices vary from 1 to 3, and metric signature is +2. The proper time along the world line $P(\tau)$ is $\tau$, the comoving time of an orthonormal unit tetrad $e^\mu$ is $t$, and $\Gamma^\mu_{\alpha\beta}$ are the Christoffel terms that in turn are related to the Riemann curvature tensor $R^\rho_{\mu\nu\sigma}$ which is the principal source of the physical deviation or acceleration in (1). For the motion of a free particle, the space is flat whereby $R^\rho_{\mu\nu\sigma} = 0$ and there is only nongravitational and rotational acceleration.

Given an observer's 4-velocity $u(\tau)=(u^o,\boldsymbol{v})$ and 4-rotation $\omega(\tau)=(\omega^o,\boldsymbol{\omega})= (\omega^o, \vec{\omega})$ along with a modified form of (1) due to Manasse & Misner (1963), the resulting equations of motion to second-order in $\boldsymbol{x}$ for a particle of mass $m$ are

$$\ddot{\boldsymbol{x}} = \ddot{\boldsymbol{x}}_{Inertial} + \ddot{\boldsymbol{x}}_{Riemann} \; , \quad (2a)$$

$$\ddot{\boldsymbol{x}}_{Inertial} = -(1+\boldsymbol{a}\cdot\boldsymbol{x})\boldsymbol{a} + 2(\boldsymbol{a}\cdot\boldsymbol{v})(1+\boldsymbol{a}\cdot\boldsymbol{x})\boldsymbol{v} + (\boldsymbol{b}\cdot\boldsymbol{x})\boldsymbol{v}$$
$$- 2(1+\boldsymbol{a}\cdot\boldsymbol{x})(\vec{\omega}\times\boldsymbol{v}) + \vec{\omega}\times(\vec{\omega}\times\boldsymbol{v}) + (\vec{\eta}\times\boldsymbol{x}) + O(\boldsymbol{x}^2)$$
$$(2b)$$

$$\ddot{x}^i_{Riemann} = \ddot{x}^i_F = -\,^F R_{oioj}\, x^j - 2\,^F R_{ikj0}\dot{x}^k x^j$$
$$- \frac{2}{3}(3\,^F R_{0kj0}\dot{x}^i\dot{x}^k +\,^F R_{ikjl}\dot{x}^k \dot{x}^l$$
$$+ \,^F R_{0kjl}\dot{x}^i\dot{x}^k \dot{x}^l\, )x^j + O(\boldsymbol{x}^2)$$
$$(2c)$$

where (2c) is not written in bold-face vector notation like in (2a,b) due to the complexity of the $R^\rho_{\mu\nu\sigma}$ terms. The symbol $\boldsymbol{a}$ is the nongravitational inertial acceleration of the observer's tetrad frame (e.g. thrust on Pioneer) as well as the classical Newtonian gravitational acceleration $a = -\nabla\Phi = \partial_j \Gamma^j_{oo}$ in the weak-field limit, and $\boldsymbol{v}$ is frame velocity. The dot-notation (e.g. for velocity $\boldsymbol{v} = \dot{\boldsymbol{x}} = \partial_\tau \boldsymbol{x}$) is used in general for differentiation with respect to $\tau$, and the subscript-superscript $F$ refers to Fermi coordinates. New terms have been introduced as $b \equiv \nabla_u a = (b^o,\boldsymbol{b})$ and $\eta \equiv \nabla_u \omega = (\eta^o,\vec{\eta})$ where $\nabla_u$ is the covariant derivative with respect to $u$. These have components $b^o=a^2$; $\boldsymbol{b} = \dot{\boldsymbol{a}} + \omega\times a$; $\eta^o=\omega\cdot a$; and $\boldsymbol{\eta}= \dot{\vec{\omega}}$. The $\boldsymbol{a}\cdot\boldsymbol{x}$ term represents Doppler redshift corrections; the $\boldsymbol{a}\cdot\boldsymbol{v}$ term special relativistic corrections; the $\boldsymbol{b}\cdot\boldsymbol{x}$ term changes in redshift corrections; and $\boldsymbol{\eta}$ angular acceleration. One must be careful in the classical limit with the signs in (2) arising from the choice of metric, because classically the Coriolis and centrifugal terms have the same sign.

To second-order in $\boldsymbol{x}$ there is no mixing of the inertial accelerations (2b) with the gravitational effects induced under Fermi transport by Riemann curvature in (2c). The respective terms in (2) have been derived, discussed, and tabulated elsewhere (Ni and Zimmermann 1978; Li and Ni 1979). In addition Li & Ni (1979) have expanded the procedure to third-order in $\boldsymbol{x}$ to illustrate the admixture of nongravitational and gravitational curvature terms that first appears in the next order of approximation. The Fermi-frame curvature terms in (2c) have also been derived by Chicone & Mashhoon (2002, 2006).

Since Pioneer is travelling at nonrelativistic velocities and accelerations, the relativistic terms $\boldsymbol{a}\cdot\boldsymbol{x}$, $\boldsymbol{a}\cdot\boldsymbol{v}$, and $\boldsymbol{b}\cdot\boldsymbol{x}$ are all negligible. Similarly, angular accelerations $\eta$ are negligible. Then (2) can be simplified to

$$\ddot{x} = -a_{Non\text{-}Grav} - \nabla\Phi - 2(\vec{\omega} \times v) + \vec{\omega} \times (\vec{\omega} \times v) + \ddot{x}_{Riemann} \quad (3a)$$

where $\Phi$ is the classical Newtonian potential. The centrifugal term $\omega \times (\omega \times v)$ can be absorbed into the gravitational potential $\Phi$, creating an effective velocity potential $\Phi_{Eff}$ in which these terms ostensibly disappear, a technique used in Galactic astronomy (Binney & Tremaine 2008). In that case, (3a) further simplifies to

$$\ddot{x} = -a_{Non\text{-}Grav} - \nabla\Phi_{Eff} - 2(\vec{\omega} \times v) + \ddot{x}_{Riemann} \quad . \quad (3b)$$

To first-order in $x$, only the first right-hand curvature term in (2c) appears in (3b). Assuming a free-falling observer ($a_{Non\text{-}Grav}=0$) in a nonrotating Fermi frame ($\omega=0$), then (3b) becomes

$$\frac{dx^j}{d\tau^2} = -\nabla\Phi_{Eff} - {}^F R^j{}_{0k0}\, x_F{}^k \quad . \quad (3c)$$

Clearly geodesic motion for an accelerating, rotating observer (Figure 1) depends upon the metric background that defines the Riemann geometry involved, with the equations of motion following from (2) and (3). Note that (3c) contains a Fermi-frame term that is observer and Riemann-tensor dependent, derived using perturbation methods. There is also another word of caution. Except for the use of covariant derivatives as the nabla operator $\nabla_u$, General Relativity is not clear on the subject of multiple interactions (e.g. as compared to quantum field theory). One must be very careful not to superpose multiple metrics when examining (3). Einstein and Straus (1945, 1946) studied the problem of embedding a Schwarzschild metric (representing the Solar System) into a different expanding cosmological metric. The method appears contrived and the situation to date is still under discussion (Gautreau 1984; Van den Bergh & Wils 1984; Balbinot et al. 1988; Senovilla & Vera 1997; Mars 1998; Bonnor 1999). The method of geodesic deviation adopted here avoids the Einstein-Straus problem since no such assumption has been made in the derivation of (3).

*3.3 Metrics for Cosmic Expansion*

For problems addressing the effects of universal expansion upon the local observer as described in (2) and (3), one can adopt the Friedmann-Lemaître (FL) models of "big bang" cosmology and introduce the metric in Robertson-Walker coordinates (RW) for an Einstein-De Sitter space

$$ds^2 = -dt^2 + a(t)(dx^2 + dy^2 + dz^2) \quad , \quad (4a)$$

or for the related metric in isotropic RW form

$$ds^2 = -dt^2 + a(t)\left[\frac{dr^2}{1-kr^2}\right] + r^2 d\Omega^2 \quad , \quad (4b)$$

where $d\Omega^2 = d\theta^2 + \sin^2\theta d\varphi^2$ is in spherical coordinates $(r, \theta, \varphi)$, $k$ is the curvature parameter, and $a = a(t)$ is the FL scale factor of expansion determinable from Einstein's field equations the once a matter distribution is specified. This standard notation $a(t)$ is not to be confused with the acceleration terms appearing in (2) and (3).

The FLRW metric in (4b) has been studied by Blome & Wilson (2005) in isotropic form for a flat accelerating $\Lambda$CDM (Cold Dark Matter) universe. The scale factor of expansion $a(t)$ was determined to be

$$a(t) = \sqrt[3]{\frac{\Omega_m}{2\Omega_\Lambda}\left(\cosh\left(\frac{t}{t_V}\right) - 1\right)} \quad , \quad (5)$$

with the approximation

$$t < t_V \quad a(t) \sim t^{2/3} \quad (6a)$$

$$t > t_V \quad a(t) \sim exp\left(\frac{t}{t_V}\right) \quad , \quad (6b)$$

where $t_V = (3\Lambda c^2)^{-1/2}$ is the timescale set by the vacuum energy represented by $\Lambda$. Here $\Omega_m$ and $\Omega_\Lambda$ are the contributions of total matter $m$ (baryonic plus dark) and $\Lambda$ to the closure parameter $\Omega_{cp}$, respectively. In a Euclidean FL universe (4b) both a decelerating and an accelerating phase occur as depicted in Figure 2. The transition between deceleration-acceleration occurs at a redshift $z_* = (2\Omega_\Lambda/\Omega_m)^{1/3} - 1$. If $\Omega_m = 0.3$ and $\Omega_\Lambda = 0.7$ this corresponds to $z_* \approx 0.7$ and the transition happens at a time $t_* = 7.5$ Gyr. Note that $t_v$ is approximately equal to $t_*$ ($t_v \approx t_*$).

For the FLRW metric (4b), the Riemann expansion term in (3c) becomes

$${}^F R^j{}_{0k0} = \frac{\ddot{a}}{a}, \quad (7a)$$

which further depends upon matter density $\rho$, pressure $p$, and $\Lambda$ as follows from the Friedmann equation

$$\frac{\ddot{a}}{a} = -\frac{4\pi G}{3}(\rho + \frac{3p}{c^2}) + \frac{\Lambda c^2}{3} \quad . \qquad (7b)$$

With reference again to Figure 2, ordinary matter dominates over the second $\Lambda$-term in (7b) at early times ($t < t_v$). Assuming $p=0$ (incoherent matter) in this case, we have

$$\frac{\ddot{a}}{a} \approx -\frac{4\pi G}{3}\rho \approx -\frac{H^2}{2} \approx 2.63 \times 10^{-36} \, s^{-2} \quad . \qquad (7c)$$

For the present accelerating era ($t > t_v$), the $\Lambda$-term dominates and we have

$$\frac{\ddot{a}}{a} = +\frac{\Lambda c^2}{3} \approx 5.2 \times 10^{-36} \, s^{-2} \quad , \qquad (7d)$$

where $H = \dot{a}/a$ is the Hubble parameter, $H_o$ is the present value of $H$ (assuming here $H_o$ = 71 km s$^{-1}$ Mpc$^{-1}$ = 2.29 x 10$^{-18}$ s$^{-1}$), the age of the FLRW universe is $t = t_o$= 13.8 Gyr, and $\Lambda$ = 1.74 x 10$^{-56}$ cm$^{-2}$. Relations (7) will be used below in the discussion of Pioneer effects.

*3.3 The Two-Body Problem in Accelerating, Rotating Coordinates*

Now we are prepared to address the effects of cosmic expansion upon local systems. The first-order cosmic expansion term in (3c) was arrived at for a test particle of negligible nonzero mass $m$ moving in the spacetime background of an expanding FLRW universe (4b). The presence of such a particle does not change the metric background, although the problem is actually nonlinear and this is not the case in general. Suppose now that one wants to create a bound state for the examination of local dynamics in a closed system such as a solar system or a satellite in Earth orbit. A second object of larger mass $M$ can now be placed in the proximity of the original particle $m$ in (3c). This is basically the two-body problem in a cosmological background which has been thoroughly studied (e.g. Cooperstock et al. 1998 and citations therein; Bonnor 1999). From Newton's law of gravitation, the force $F$ acting on the mass $m$ is $F = m\ddot{x} = -m\nabla\Phi$ where $\Phi$ is Newton's scalar gravitational potential created by $M$. From this relation follow all of the results of classical Keplerian mechanics. However, (3c) is actually stating that there is a universal change to Newton's law $F = m\ddot{x}$ introduced by an accelerating cosmic expansion ($\ddot{a} \neq 0$). The fundamental question is whether or not cosmological perturbations of local systems (CPLS) as in (3c) really happen, or whether CPLS is in fact a measurable or observable quantity.

When $m$ is captured in orbit about the larger mass $M$, the result is a Keplerian bound state whose orbital eccentricity is $e<1$. For the sake of simplicity, the two-body equation of motion (3c) for a circular orbit ($e=0$) of radius $r$ is

$$\ddot{\vec{r}} = -\frac{GM}{r^2}\hat{r} + \frac{\ddot{a}}{a}\vec{r} \quad , \qquad (8a)$$

where $\hat{r}$ is a radial unit vector. Cooperstock et al. (1998) argue due to the time-dependence of $\ddot{a}/a$ in (3c) and (8) that there exists a time-dependent potential well in which there is a fractional change in orbital radius over cosmic time. Adkins et al. (2007), on the other hand, disagree and conclude that the answer is negative because one cannot place a large mass $M$ on the background (4b) without solving the nonlinear field equations involved. They neglect, however, to show the solution themselves. It is elementary that a bound state of $M>>m$ can be physically taken into consideration where $(M+m)<<<M_U$ and $M_U$ is the mass of the universe that creates metric (4b). The question they ask themselves, are there measurable CPLS interactions, is not actually answered.

Within the framework of Fermi normal coordinates developed here, the influence of the $\ddot{a}/a$ term in (8a) on local dynamics can be expressed in terms of the deceleration parameter $q = -\ddot{a}a/\dot{a}^2$, where $q \to q_o$ represents the value of $q$ today. A simple calculation shows that $\ddot{a}/a = -qH^2$ and (8a) becomes

$$\ddot{\vec{r}} = -\frac{GM}{r^2}\hat{r} - q_o H_o^2 \vec{r} \quad , \qquad (8b)$$

which in a Newtonian sense can be understood as a repulsive force. The deceleration parameter can also be expressed as $q_o = (½\Omega_m - \Omega_\Lambda)$ in terms of the density parameter for matter ($\Omega_m$) and the cosmological constant ($\Omega_\Lambda$). Several authors have speculated about $cH_o$ (dimensionally an acceleration) as an anomalous acceleration for modifying Newton's law (Milgrom 2002) or for pointing out its numerical coincidence with the Pioneer deceleration $a_p$ = 8.74x10$^{-8}$ cm s$^{-1}$ (Yi 2004). $cH_o$ obviously does not follow from (8) and its connection with CPLS appears just that, a coincidence. In fact, FLRW cosmology (8) has the wrong sign for the Pioneer effect if the Universe is accelerating. The correct sign exists only if we are living in a decelerating Universe or one with non-accelerated expansion ($\Lambda<0$).

These discussions illustrate several aspects of the CPLS issue with respect to bound states in the two-body problem.

## 4. The Rotating Gödel Universe

The focus in §3 was cosmic expansion while the subject will now change to cosmic rotation. Although the possibility of a rotating universe was addressed by Lanczos (1924), it was Gödel (1949) who changed the cosmic landscape of General Relativity and cosmology with an important new solution to Einstein's field equations.

*4.1 Original Gödel Solution*

The metric discovered by Gödel is

$$ds^2 = -dt^2 - 2\sqrt{2}U(x)dtdy + dx^2 - U^2(x)dy^2 + dz^2 \qquad (9)$$

where the mixing term $dtdy$ is off-diagonal and is the source of the rotation, a situation that happens with the Kerr metric in astrophysics. Here $U(x) = exp(\sqrt{2}U\Omega x)$ is a rotation potential for the universal rotation or vorticity $\Omega$. Originally in this model where matter is described as dust with energy density $\rho$, the metric requires a cosmological constant $-\Lambda$ and the latter is directly related to the rotation as $\Omega^2 = -\Lambda = 4\pi G\rho$. Hence, $\Lambda$ and $\Omega$ are commingled if this solution exists, seeming to indicate that a vacuum energy density is necessary in order to create the intrinsic vorticity of this solution. Gödel's metric also corresponds to a rotating model universe with $\Lambda = 0$, $p = \rho$, and $\Omega^2 = 8\pi G\rho$ where $p$ is pressure (Ciufolini & Wheeler 1995).

Since the metric in (9) involves a uniform rotation of matter relative to what is called the compass of inertia, rotational Coriolis and centrifugal acceleration terms must be induced in the equations of motion for an object on a trajectory traversing such a world model. An early investigation of these explicit terms was given by Silk (1966), derived using perturbation methods.

Obviously there exists a speed-of-light circle in a rotating universe beyond which metric (9) loses its physical meaning, a pathology that is discussed in Hawking & Ellis (1973). This and the related problem that there exist closed time-like curves (CTCs) in (9) have been addressed in a long series of papers by Ozsváth & Schücking (2001, 1997, 1969) that include making the Gödel universe finite to avert the pathologies. Bonnor, Santos, & MacCallum (1998) have also studied this feature of (9).

*4.2 Cosmic Rotation in Accelerating, Rotating Coordinates*

Patterned after the procedure of §3.2, one needs the equations of motion in order to assess the behavior of an observer in such a universe. The universal rotation introduced by Gödel results in only the following nonzero Riemann curvature terms for metric (9) that appear in (2c) (Chicone & Mashhoon 2006)

$$R_{0101} = \Omega^2, \qquad R_{0202} = \Omega^2 U^2,$$
$$R_{0112} = -\sqrt{2}\Omega^2 U, \qquad R_{1212} = 3\Omega^2 U^2, \qquad (10)$$

due to symmetries in the Riemann tensor. One can then show that the nonzero components of their projection into the Fermi frame can be found from

$$^F R_{0101} = {^F R_{0202}} = {^F R_{1212}} = \Omega^2, \qquad (11)$$

again from the symmetries involved. By virtue of (11), the Gödel curvature terms in (2c) now become

$$\ddot{x}^i_F = -\Omega^2 \left\{ \left[ 1 - \delta_{3i} + \frac{2}{3}\delta_{1i}(\dot{x}^2)^2 \right] x^i + \frac{2}{3}\gamma x^o \dot{x}^i \right\} \qquad (12)$$

where $\gamma = (\dot{x}^1)^2 + (\dot{x}^2)^2$ and $\delta_{ij}$ is the standard delta function. Considering the case with $\omega \rightarrow \Omega$ in (3a) and using (12), one finally obtains the following equations of motion

$$\ddot{x}^i = -a^i - \nabla^i \Phi - 2\varepsilon_{ijk}\Omega^j v^k - \varepsilon_{ijk}\Omega^j \varepsilon^k_{lm}\Omega^l v^m$$
$$- \Omega^2 \left\{ \left[ 1 - \delta_{3i} + \frac{2}{3}\delta_{1i}(\dot{x}^2)^2 \right] x^i + \frac{2}{3}\gamma x^o \dot{x}^i \right\} \qquad (13)$$

for an accelerating, rotating observer to second-order in $x$ for the Gödel universe. The expanding cosmic term $\ddot{a}/a$ in (8) has vanished since there is no accelerating expansion ($\ddot{a}=0$) in Gödel's case and the perturbation in (13) is due to $\Omega$-terms induced by cosmic rotation. These are respectively the Coriolis effect and centrifugal force term as well as the Riemann curvature terms produced by (11).

*4.3 Gödel-Type Rotating Models With Expansion*

Metric (9) is not an expanding universe. Gödel (1950) himself considered expanding solutions with rotation which have since inspired a whole new field of cosmology known as Bianchi universes (Ellis 2000; Bergamini et al. 1997) as well as development of the Raychaudhuri (1955) equation relating expansion $\Theta$ (of a fluid), rotation or vorticity $\Omega$, and shear $\sigma$. A considerable number of exact and approximate models with both expansion and rotation have since appeared (Obukhov 2000). The subject of universal rotation is often discussed in terms of Petrov and Bianchi types (Stephani et al. 2003).

## 5. Consequences of Astrophysical Limits

For the rotation of the Solar System about the Galactic Center $\omega_{SS-GC}$, the Coriolis term in (3a,b) is $|a_c| \sim 2\omega_{SS-GC} v \approx 2 \times 10^{-9}$ cm s$^{-2}$, which is a number resembling that of the Pioneer anomaly $a_p = \sim 8.74 \times 10^{-8}$ cm s$^{-2}$ but is two orders of magnitude too small to explain the effect. In the present study, on the other hand, we are concerned with the $\Omega$ and $\Omega^2$ terms in (13). Since $\Omega^2 \sim \Lambda$ we can adopt the value of $\Lambda = 1.74 \times 10^{-56}$ cm$^{-2}$ from the accelerating FLRW model in §3.3 (Blome & Wilson 2005) as an estimate to establish a limit on rotation $\Omega$. This gives the approximation $\Omega \sim 10^{-28}$ s$^{-1}$. A similar comparison using $\Omega^2 = 4\pi G\rho$ if $\Lambda = 0$ gives the result $\Omega \sim 4 \times 10^{-18}$ s$^{-1}$.

However, there are additional rotational limits that follow from experimental measurements of anisotropies in the cosmic microwave background radiation (CMBR), number counts, and polarization rotation of electromagnetic radiation propagating in (9). Using the upper limit established by the Cosmic Background Explorer (COBE) for $(\Omega/H)_o \leq 10^{-6}$ on the rotational vorticity in the current epoch (Bunn et al. 1996; Kogut et al. 1997), a value of $H_o = 71$ km s$^{-1}$ Mpc$^{-1}$ = 2.3 x 10$^{-18}$ s$^{-1}$ for the Hubble constant gives a limit $\Omega \leq 2.3 \times 10^{-24}$ s$^{-1}$ although a more recent value $H_o = 73$ km s$^{-1}$ Mpc$^{-1}$ = 2.4 x 10$^{-18}$ s$^{-1}$ (Spergel et al. 2007) has also been given. There is a similar constraint for shear, namely $(\sigma/H)_o \leq 3 \times 10^{-9}$, and in fact there exists a direct relationship between $(\Omega/H)_o$ and $(\sigma/H)_o$ (Bunn et al. 1996, Ref. 6). In order to avert these limits, Obukhov (2000) has argued that it is actually shear $\sigma$, not vorticity $\Omega$, that is the source of CMBR anisotropy. Since Bunn et al. have shown that $\sigma$ and $\Omega$ are directly related for a Bianchi model of type VII$_h$, the argument is not particularly compelling.

Because the $(\Omega/H)_o$ limit results in $\Omega^2 \sim 10^{-48}$ s$^{-2}$, the last two rotational vorticity terms on the right-hand-side of (13) are negligible compared to the Coriolis term. For an escape velocity from the Solar System of the Pioneer spacecraft $v_{Pioneer} = 11 \times 10^6$ cm s$^{-1}$, the Coriolis term is $|a_c| \sim 2 \Omega_{Gödel} v_{Pioneer} \approx 2 \times 10^{-18}$ cm s$^{-2}$. This is ten orders of magnitude smaller than $a_p$. Although it contributes, a universal rotation $\Omega_{Gödel}$ is not the cause of the Pioneer effect.

## 6. Frame-Dragging Effects

Surprising results arising from Einstein's then-new theory of gravitation were found by De Sitter (1916) and Thirring (1918) with planetary applications by Lense & Thirring (1918), the latter now commonly referred to as frame-dragging. Upon calculating rotational terms appearing in the Keplerian orbit equations of motion for General Relativity, they obtained Coriolis and centrifugal accelerations like those appearing in (3a). The appearance of such terms was actually nothing new. What was significant, however, was where they were coming from – the Riemann curvature tensor and relativistic nature of Einstein's theory. The question here is how and if frame-dragging has something to do with the Pioneer anomaly.

There exist many discussions of Lense-Thirring (1918) frame-dragging (e.g. Ciufolini 1986; Zeldovich & Novikov 1971) for a test particle in orbit with semi-major axis $a$ and eccentricity $e$ about a central spherical body having angular momentum $\mathbf{J}$. In the weak-field, slow-motion approximation, the line of nodes is dragged in the sense of rotation at a rate $\dot{\mathbf{\Omega}}$, where

$$\dot{\mathbf{\Omega}} = [2/a^3(1-e^2)^{3/2}]\mathbf{J} \quad . \tag{14}$$

For unbound states such as parabolic and hyperbolic trajectories with eccentricity $e \geq 1$, (14) contains a singularity or an imaginary root and is no longer a valid physical equation. However, no one appears to have investigated frame-dragging for hyperbolic trajectories as is the case for Pioneer – although the tidal terms for stellar encounters involving parabolic ($e = 1$) equatorial orbits have been published (Ishii et al. 2005) and frame-dragging has been addressed for FLRW backgrounds in cosmology (Schmid 2009).

## 7. The Virial Theorem and Hyperbolic Trajectories

Earth fly-by's and the Pioneer anomaly both involve close orbital encounters at small impact parameters resulting in hyperbolic trajectories with respect to a planet in the Solar System. They both involve transitions from bound to unbound states in one fashion or another, derived from patched-conic techniques used in astrodynamics. We will now show that energy cannot be conserved in such a procedure by virtue of the cosmological virial theorem.

The virial theorem provides a general relation between the time-averaged total kinetic energy $T=<T>$ and potential energy $U=<U>$ such that the virial energy $2T+U$ is zero: $2T+U=0$. It applies for a self-gravitating system of equal-mass objects (stars, galaxies, etc.) in stable equilibrium and is used to examine the stability of galactic clusters believed to have negative total energy $E$, the classical definition of a bound state. Briefly, a stable bound-state system's potential energy must equal its kinetic energy within a factor of two. In the general case of finite classical motion, not necessarily along a bound orbit, the virial theorem $2T+U=0$ is valid for ergodic quantities $f$ averaged over a large time interval as $<f>_t$. Quantities can also be averaged over other parameters such as eccentric anomaly $\mathcal{E}$ as $<f>_{\mathcal{E}}$ and true anomaly $\varphi$ as $<f>_\varphi$. Examples of such mean values are given in the literature (Serafin 1980; Szebehely 1989).

The Layzer-Irvine equation (Layzer 1963; Irvine 1961) is an extension of the virial theorem to systems that interact with an expanding cosmic environment ($\dot{a}/a > 0$). It relates the total system energy $E = T+U$ with the virial energy $2T+U$ as follows:

$$\frac{d}{dt}(T+U) + \frac{\dot{a}}{a}(2T+U) = 0 \quad , \qquad (15)$$

where $\dot{a}$ is the scale factor of expansion appearing in the Hubble parameter $H = \dot{a}/a$. Note that the expansion term in (15) is similar to the CPLS-type interaction in (8) found by Cooperstock et al. (1998) and this study. Both (15) and (8) are consistent with the results of Anderson (1995).

Expansion causes the energy of a system *not* in virial equilibrium to change because from (15)

$$-\frac{d}{dt}(T+U) = \frac{\dot{a}}{a}(2T+U) = 0 \qquad (16)$$

which can happen *only* when the virial energy is zero, $2T+U = 0$. Otherwise total energy $E$ cannot be conserved in accordance with the left-hand side of (16).

### 7.1 Space Astrodynamics

For a general Keplerian orbit, (16) reads:

$$-\frac{d}{dt}(-\frac{GMm}{2r_a}) = \frac{\dot{a}}{a}\left\{m(V_r^2+V_\phi^2)-\frac{GMm}{r}\right\} \quad (17)$$

for velocity $V$, semi-major axis $r_a$, eccentricity $e$, and true anomaly $\varphi$. This is also (Szebehely 1989)

$$-\frac{d}{dt}(-\frac{GM}{2r_a}) = \frac{\dot{a}}{a}\left\{\frac{GMe(e+cos\phi)}{r_a(1-e^2)}\right\} \qquad (18)$$

while (17) becomes

$$-\frac{d}{dt}(-\frac{GM}{2r_a}) = \frac{\dot{a}}{a}\left\{\frac{GM}{\langle r \rangle}-\frac{GM}{r_a}\right\} \qquad (19)$$

when the *vis viva* equation is used for velocity $V$ (with the method of virial averaging $<r>$ also indicated).

### 7.2 Bound-State Orbit (Circular, e=0)

For a bound-state Keplerian orbit with zero eccentricity, (18) becomes

$$-\frac{d}{dt}(-\frac{GM}{2r_a}) = \frac{\dot{a}}{a}\{0\} = 0 \quad , \qquad (20)$$

and it follows from the left-hand side of (16) that total energy $E$ is conserved.

### 7.3 Bound-State Orbit (Elliptical, 0<e<1)

For a bound-state Keplerian elliptical orbit, $2T+U=0$ is valid because it is a bound system like the circular case §7.1. Calling attention to (19), $<r>$ is the virial average with respect to the eccentric anomaly $\varphi$ or $<r>_\varphi$, and this becomes $<r>_\varphi = r_a$ (Szebehely 1989). (19) is now

$$-\frac{d}{dt}(-\frac{GM}{2r_a}) = \frac{\dot{a}}{a}\left\{\frac{GM}{r_a}-\frac{GM}{r_a}\right\}=\frac{\dot{a}}{a}\{0\}=0. \qquad (21)$$

It follows from the left-hand side of (16) that total energy $E$ is conserved.

### 7.4 Unbound Orbit (Hyperbolic Trajectory, e>1)

For an unbound Keplerian orbit on a hyperbolic trajectory, (18) becomes

$$-\frac{d}{dt}(-\frac{GM}{2|r_H|}) = -\frac{\dot{a}}{a}\left\{\frac{GMe(e+cos\phi)}{|r_H|(e^2-1)}\right\} \neq 0 \qquad (22)$$

where $r_H$ is the semi-major axis for the hyperbolic case and $r_H < 0$. It is obvious that the total energy $E=T+U$ is not conserved in (22) due to the left-hand side of (16). Note that one cannot use the argument in §7.3 for hyperbolic trajectories because $r$ has no periodic motion and hence the method of virial averaging is not applicable. The same argument applies to the parabolic case ($e=1$).

The above discussion shows that hyperbolic trajectories in (22) such as that for Pioneer are influenced by cosmic expansion $H = \dot{a}/a$ while closed conical orbits like those in (20) and (21) are not. The source of the nonconservation of energy is the cosmic perturbation of local dynamics caused by the expansion term in (15). The Layzer-Irvine equation, then, confirms the argument of Anderson (1995) that cosmic expansion couples to escape orbits while it does not affect bound orbits.

## 8. Conclusions

We have derived the equation of motion for an accelerated, rotating observer in a Gödel universe, and have shown that the universal cosmic rotation or vorticity $\Omega$ contributes to but cannot account for the Pioneer anomaly. This limitation is due to the measured anisotropy bounds found in the cosmic microwave background that constrain shear $\sigma$ and vorticity. Although the Gödel metric is not a realistic case because the present-day universe is known to be expanding, it nevertheless provides a rigorous mathematical model for characterizing cosmic rotation in order to parameterize vorticity $\Omega$ and study how it relates to the Pioneer effect. Future work will expand this result to Bianchi universes in general with the aid of the Raychaudhuri equation (1955). The Lense-Thirring effect likewise cannot account for the Pioneer anomaly because it is only applicable for bound-state orbits with eccentricity $e < 1$. Pioneer, on the other hand, experiences the anomaly on a hyperbolic trajectory with positive energy. By virtue of the virial theorem, the nonconservation of energy for a Pioneer-type observer while changing from bound-state elliptical trajectories to hyperbolic ones has been pointed out. This result illustrates how the physics of such trajectory techniques for interplanetary fly-by's is still not understood.

## Acknowledgements

The authors would like to thank an anonymous referee for several helpful comments that improved the quality of the final manuscript. We are also grateful to the 37th COSPAR Scientific Assembly in Montreal for their invitation to a plenary session where a preliminary version of these results was presented orally July 17, 2008.

## References

Adkins G.S., McDonnell J., Fell R.N. Cosmological perturbations on local systems. Phys. Rev. D 75, 064011, 2007.

Anderson J.D., Campbell J.K., Ekelund J.E., Ellis J., Jordan J.F. Anomalous orbital-energy changes observed during spacecraft flybys of Earth. Phys. Rev. Lett. 100, 091102, 2008.

Anderson, J.D., Laing P.D., Lau E.L., Liu A.S., Nieto M.M., Turyshev S.G. Study of the anomalous acceleration of Pioneer 10 and 11. Phys. Rev. D 65, 082004, 2002.

Anderson J.D., Laing P.D., Lau E.L., Liu A.S., Nieto M.M., Turyshev S.G. Indication, from Pioneer 10/11, Galileo, and Ulysses data, of an apparent anomalous, weak, long-range acceleration. Phys. Rev. Lett. 81, 2858-2861, 1998.

Anderson J.L. Multiparticle dynamics in an expanding universe. Phys. Rev. Lett. 75, 3602-3604, 1995.

Balbinot R., Bergamini R., Comastri A. Solution of the Einstein-Straus problem with a Λ–term. Phys. Rev. D 38, 2415-2418, 1988.

Bergamini R., Sedici P., Verrocchio P. Inflation for Bianchi type IX models. Phys. Rev. D 55, 1896-1900, 1997.

Binney J., Tremaine S. Galactic Dynamics, §3.3.2, second ed. Princeton Univ. Press: Princeton, 2008.

Blome H.-J., Wilson T.L. The quantum temperature of accelerating cosmological models of an entangled universe. Adv. Space Res. 35, 111-115, 2005.

Bonnor W.B. Size of a hydrogen atom in the expanding universe. Class. Quant. Grav.16, 1313-1321, 1999.

Bonnor W.B. The cosmic expansion and local dynamics. Mon. Not. Roy. Astron. Soc.282, 1467-1469, 1996.

Bonnor W.B. The cosmic expansion and local dynamics. Ap. J. 316, 49-51, 1987.

Bonnor W.B., Santos N.O., MacCallum M.A.H. An exterior for the Gödel spacetime. Class. Quant. Grav. 15, 357-366, 1998.

Bunn E.F., Ferreira P.G., Silk J. How anisotropic is our universe? Phys. Rev. Lett. 77, 2883-2886, 1996.

Ciufolini I. Measurement of the Lense-Thirring drag on high-altitude, laser-ranged artificial satellites. Phys. Rev. Lett. 56, 278-281, 1986.

Ciufolini I., Wheeler J. Gravitation and inertia. Princeton University Press, Princeton, 1995.

Chicone C., Mashhoon B. Explicit Fermi coordinates and tidal dynamics in de Sitter and Gödel spacetimes. Phys. Rev. D 74, 064019, 2006.

Chicone C., Mashhoon B. The generalized Jacobi equation. Class. Quantum Grav. 19, 4231-4248, 2002.

Cooperstock F.I., Faraoni V., Vollick D.N. The influence of the cosmological expansion on local systems. Ap. J. 503, 61-66, 1998.

De Sitter W. On Einstein's theory of gravitation, and its astronomical consequences. Mon. Not. Roy. Astron. Soc. 77, 155-184, 1916. Erratum, ibid, 481, 1916.

Einstein A., Straus E.G. The influence of the expansion of space on the gravitational fields surrounding the individual stars. Rev. Mod. Phys. 17, 120-124, 1945.

Einstein A., Straus E.G. Corrections and additonal remarks to our paper: The influence of the expansion of space on the gravitational fields surrounding the individual stars. Rev. Mod. Phys. 18, 148-149, 1946.

Ellis G.F.R. Editor's note on Kurt Gödel. Gen. Rel. Grav. 32, 1399-1408, 2000.

Fermi E. Sopra i fenomeni che avvengono in vicinanza di una linea araria. Atti Acad. Naz. Lincei Rend. Cl. Sci. Fiz. Mat. Nat. 31(I), 21-23, 51-52, 101-103, 1922.

Gautreau R. Imbedding a Schwarzschild mass into cosmology. Phys. Rev. D 29, 198-206, 1984.

Gödel K. An example of a new type of cosmological solutions of Einstein's field equations of gravitation. Rev. Mod. Phys. 21, 447-450, 1949.

Gödel K. Rotating universes in general relativity theory. In: Graves L.M., Hille E., Smith P.A., Zariski O. (Eds.), Proc. Int'l. Congress of Mathematicians (Cambridge, USA, 1950), Amer. Math. Soc.: Providence, RI, Vol. 1, 175-181, 1952. (Reprinted in Gen. Rel. Grav. 32, 1419-1427, 2000.)

Hawking S., Ellis G.F.R. The large-scale structure of space-time. Cambridge University Press, Cambridge, 1973, §5.7.

Irvine, D. Local irregularities in a universe satisfying the cosmological principle. Thesis, Harvard University, 1961.

Ishii M., Shibata M., Mino Y. Black hole tidal problem in Fermi normal coordinates. Phys. Rev. D 71, 044017, 2005.

Kogut A., Hinshaw G., Banday A.J. Limits to global rotation and shear from the COBE DMR four-year sky maps. Phys. Rev. D 55, 1901-1905, 1997.

Lämmerzahl C., Preuss O., Dittus H. Is the physics within the Solar System really understood? In Dittus, H., Lämmerzahl, C., Turyshev, S.G. (Eds.), *Lasers, Clocks, and Drag-free Control - Exploration of Relativistic Gravity in Space*, Springer-Verlag, Berlin Heidelberg, 75-101, 2008.

Lanczos K. Über eine stationäre Kosmologies im Sinne der Einsteinschen Gravitationstheorie. Zeits. f. Phys. 21, 73-110, 1924; On a stationary cosmology in the sense of Einstein's theory of gravitation. Gen. Rel. Grav. 29, 363-399, 1997.

Layzer D. A preface to cosmogony. I. The energy equation and the virial theorem for cosmic distributions. Ap. J. 138, 174-184, 1963.

Lense J., Thirring H. Über den Einfluss der Eigenrotation der Zentralkörper auf die Bewegung der Planeten und Monde nach der Einsteinschen Gravitationstheorie, Physik. Zeitschr. 19, 156-163, 1918. English translation in Mashoon *et al.* 1984.

Li W.-Q., Ni W.-T. Coupled inertial and gravitational effects in the proper reference frame of an accelerated, rotating observer. J. Math. Phys. 20, 1473-1480, 1979.

Manasse F.K., Misner C.W. Fermi normal coordinates and some basic concepts in differential geometry. J. Math. Phys. 4, 735-745, 1963.

Mars M. Axially symmetric Einstein-Straus models. Phys. Rev. D 57. 3389-3400, 1998.

Mashhoon B., Hehl F.W., Theiss D.S. On the gravitational effects of rotating masses: The Thirring-Lense papers. Gen. Rel. Grav. 16, 711-750, 1984.

McVittie G.C. The mass-particle in an expanding universe. Mot. Not. R. Aston. Soc. 93, 325-339, 1933.

Mena F.C., Tavakol R., Vera R. Generalization of the Einstein-Straus model to anisotropic settings. Phys. Rev. D 66, 044004, 2002.

Milgrom M. Do modified Newtonian dynamics follow from the cold dark matter paradigm? Ap. J. 571, L81-L83, 2002.

Ni W.-T., Zimmermann M. Inertial and gravitational effects in the proper reference frame of an accelerated, rotating observer. Phys. Rev. D 17, 1473-1476, 1978.

Nieto M.M. New Horizons and the onset of the Pioneer anomaly. Phys. Lett. B 659, 483-485, 2008.

Nieto M.M. Analytic gravitational-force calculations for models of the Kuiper Belt, with application to the Pioneer anomaly. Phys. Rev. D 72, 083004, 2005.

Nieto M.M., Turyshev S.G. Finding the origin of the Pioneer anomaly. Class. Quant. Grav. 21, 4005-4023, 2004.

Obukhov Y.N. On physical foundations and observational effects of cosmic rotation. In: Scherfner M., Chrobok T., Shefaat M. (Eds.), Colloquium on cosmic rotation, Wissenschaft und Technik Verlag, Berlin, 23-96, 2000. ArXiv astro-phys/0008106.

Ozsváth I. The finite rotating universe revisited. Class. Quantum Grav. 14, A291-A297, 1997.

Ozsváth I., Schücking E. The finite rotating universe. Ann. Phys. 55, 166-204, 1969.

Ozsváth I., Schücking E. Approaches to Gödel's rotating universe. Class. Quantum Grav. 18, 2243-2252, 2001.

Raychaudhuri A. Relativistic cosmology, I. Phys. Rev. 98, 1123-1126, 1955.

Schmid C. Mach's principle: Exact frame-dragging via gravitomagnetism in perturbed Friedmann-Robertson-Walker universes with $K = (\pm 1, 0)$. Phys. Rev. D 79, 064007, 2009.

Senovilla J.M.M., Vera R. Impossibility of the cylindrically symmetric Einstein-Straus model. Phys. Rev. Lett. 78, 2284-2287, 1997.

Serafin R.A. The mean anomaly in elliptic motion as random variable. Celestial Mech. 21, 351-356, 1980.

Silk J. Local irregularities in a Gödel universe. Ap. J. 143, 683-699, 1966.

Stephani H., Kramer D., MacCallum M., Hoenselaers S., Herlt E. Exact solutions of Einstein's field equations. Second edition, Cambridge University Press, Cambridge 2003.

Spergel D.N. et al. Three-year *Wilkinson Microwave Anisotropy Probe (WMAP)* observations: Implications for cosmology. Ap. J. Suppl. 170, 377-408, 2007.

Szebehely V.G. Adventures in celestial mechanics. University of Texas Press, Austin, 1989, pp. 53-54.

Thirring H. Über die Wirkung rotierender ferner Massen in der Einsteinschen Gravitationstheorie. Physik. Zeitschr. 19, 33-39, 1918. English translation in Mashoon *et al.* 1984.

Turyshev S.G., Toth V.T., Kellogg L.R., Lau E.L, Lee K.J. The study of the Pioneer anomaly: new data and objectives for new investigation. Int'l. J. Phys. D 15, 1055, 2006.

Turyshev S.G., Nieto M.M., Anderson J.D. Study of the Pioneer anomaly: a problem set. Am. J. Phys. 73, 1033-1044, 2005.

Van den Bergh N., Wils P. Imbedding a Schwarzschild mass into cosmology. Phys. Rev. D 29, 3002-3003, 1984.

Yi Y.G. On the indication from Pioneer 10/11 data of an apparent anomalous, weak, long-range acceleration. Gen. Rel. Grav. 36, 875-881, 2004.

Zeldovich Ya.B., Novikov I.D. Relativistic astrophysics: Stars and relativity. Vol. 1. University of Chicago Press, Chicago, 1971, §1.10.

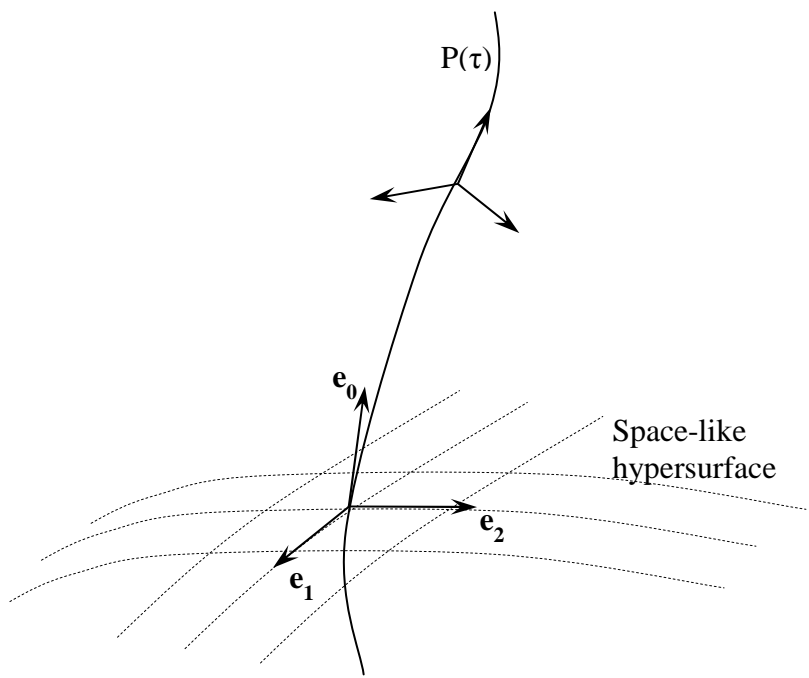

*Figure 1. An observer's rotating Fermi frame is depicted along a world line P(τ) with proper time τ, used to derive Equations (2)-(3). $e_3$ is hidden in the space-like hypersurface.*

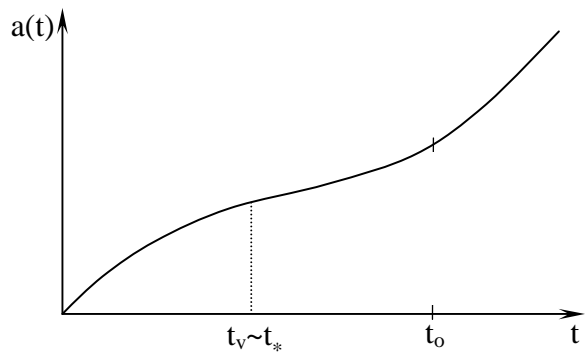

*Figure 2. Friedmann-Lemaitre cosmic expansion dynamics for a(t), illustrating the regions of Equations (7c) and (7d).*